\documentclass[12pt]{iopart}
\usepackage{iopams}
\usepackage{amssymb,latexsym}
\usepackage{amstext,amsthm}

\usepackage{amsgen,amsfonts,amsbsy}
\usepackage{epsfig,graphicx}
\usepackage{amscd}
\usepackage{lscape}
\newcommand{\Real}{\mathbb R}

        \newcommand{\beq}{\begin{eqnarray}}
        \newcommand{\eeq}{\end{eqnarray}}
        
        \newcommand{\ket}[1]{$ |{#1}\rangle$}
        
        \newcommand{\mbra}[1]{\langle {#1}|}
        \newcommand{\mket}[1]{ |{#1}\rangle}

        \newcommand{\sks}[2]{\langle {#1}|{#2}\rangle }

\begin{document}

\title[Coherent states on the circle]{Quantizations
on the circle and coherent states}
\author{G Chadzitaskos, P Luft and J Tolar}
\address{Department of Physics\\
Faculty of Nuclear Sciences and Physical Engineering   \\
Czech Technical University in Prague\\ B\v rehov\'a 7,  CZ - 115 19
Prague, Czech Republic}

\eads{\mailto{jiri.tolar@fjfi.cvut.cz},
\mailto{goce.chadzitaskos@fjfi.cvut.cz}}

\begin{abstract}
We present a possible construction of coherent states on the unit
circle as configuration space. Our approach is based on Borel
quantizations on $S^{1}$ including the Aharonov-Bohm type quantum
description. The coherent states are constructed by Perelomov's
method as group related coherent states generated by Weyl operators
on the quantum phase space $\mathbb{Z} \times S^{1}$. Because of the
duality of canonical coordinates and momenta, i.e. the angular
variable and the integers, this formulation can also be interpreted
as coherent states over an infinite periodic chain. For the
construction we use the analogy with our quantization and coherent
states over a finite periodic chain where the quantum phase space
was $\mathbb{Z}_{M} \times \mathbb{Z}_{M}$. The coherent states
constructed in this work are shown to satisfy the resolution of
unity. To compare them with canonical coherent states, also some of
their further properties are studied demonstrating similarities as
well as substantial differences.
\end{abstract}

 \pacs{03.65.-w, 03.65.Fd, 02.20.-a}
 \submitto{J. Phys. A: Math. Theor.}
 %\maketitle

%\noindent Keywords:  Borel quantizations, Aharonov-Bohm type
%quantizations on the circle, group related coherent states, Weyl
%operators, quantum phase space

\section{Introduction}

Quantum description of a particle on a circle is one of the basic
problems tackled in quantum mechanics from its beginnings. The
reason is that rotational motion represents integral part of
important quantum models of atoms, molecules or atomic nuclei.

Coherent states belong to the most useful tools in many applications
of quantum physics \cite{AAG}. They find numerous applications in
quantum optics, quantum field theory, condensed matter physics,
atomic physics, etc. There exist various definitions and approaches
to the coherent states which depend on author and application
\cite{Kla}. Our main reference is \cite{Per1} since we base our
construction of coherent states on the circle on Perelomov's idea of
group-related coherent states, contrary to the overwhelming opinion
that this method is not suitable for this purpose.

The problem of coherent states on the circle was investigated by S.
de Bi\`evre \cite{SDB}, and also by J. A. Gonz\'alez and M. A. del
Olmo \cite{dOG}. They used the Weil-Brezin-Zak transform
$\mathcal{T}$
\begin{equation}\label{}
  \mathcal{T}:L^{2}(\mathbb{R}) \rightarrow
  L^{2}(S^{1} \times S^{1\ast})
\end{equation}
defined by
\begin{equation}\label{}
  (\mathcal{T}\psi)(q,k) :=
  \sum_{n=-\infty}^{\infty}e^{iank}\psi(q-na),
\end{equation}
\begin{equation}\label{}
\nonumber q \in S^{1} = [0,a), \quad k \in S^{1\ast} =
[0,\frac{2\pi}{a}), \quad \psi \in L^{2}(\mathbb{R}).
\end{equation}
Applying this transform to canonical coherent states on the real
line they obtained a family of coherent states on the circle labeled
by the the variables of the cylinder $\mathbb{R} \times S^{1}$. It
should be noted that a very detailed study of deformation
quantization on the cylinder as classical phase space was recently
given in \cite{GdOT}. The results confirm that in quantum theory the
quantum phase space $\mathbb{Z} \times S^{1}$ is involved, which is
central point in this investigation.

A different approach was employed by C. J. Isham and J. R. Klauder
\cite{IaK}. They constructed coherent states on the circle by using
representations of the Euclidean group $E(2)$. Being the semi-direct
product of groups $\mathbb{R}^{2}$ and $SO(2)$, it involves the
action on the angular variable. However, they observed that there
does not exist an irreducible representation of $E(2)$ such that the
resolution of unity holds. Therefore they considered only reducible
representations and extended the method to the case of the
$n$-dimensional sphere. Other definitions are based on the Lie
algebra of $E(2)$ \cite{KR08, KR96} (see also \cite{Ni}).

We adopt a general method of quantization on configuration manifolds
called Borel quantization \cite{DST01}. According to \cite{DST01}
inequivalent quantum Borel kinematics on a configuration manifold
$M$ are classified by elements of the set
$$ H^{2}(M, \mathbb{Z}) \times H^{1}(M, U(1)).
$$
Since $\mbox{dim}\, S^1=1$, this implies that the set of
inequivalent Borel quantizations is isomorphic to the set $H^{1}(M,
U(1))$ of characters of the first homology group
$H_{1}(S^1)=\mathbb{Z}$ (or equivalently, of the fundamental group
$\pi_{1}(S^1)=\mathbb{Z}$). Distinct characters
\begin{equation}\label{}
 L^{\theta}: \mathbb{Z}\rightarrow U(1):n \mapsto e^{2\pi i\theta n}
\end{equation}
are parametrized by $\theta \in [0,1)$.

In general a family of coherent states consists of vectors
$|x\rangle$ of some separable Hilbert space $\mathcal{H}$, labeled
by some parameter $x \in \mathcal{X}$. Crucial property common to
all families of coherent states is the \textit{resolution of unity}:
there exists a positive measure $d\mu(x)$ on $\mathcal{X}$ such that
\begin{equation}\label{rroouux}
  \int_{\mathcal{X}}|x\rangle\langle x|d\mu(x)=\hat{I},
\end{equation}
where $\hat{I}$ is  the unit operator. The existence of the
resolution of unity (\ref{rroouux}) has to be verified for each
family of coherent states. Then, if verified, the resolution of
unity (\ref{rroouux}) entails the completeness of the set of
coherent states, i.e. that the closed linear span of the family of
coherent states $\{|x\rangle\ \; | \, x \in \mathcal{X}\}$ is the
entire Hilbert space $\mathcal{H}$. This property means that any
vector in $\mathcal{H}$ may be represented as a linear superposition
of coherent states.

Our approach to the construction of coherent states will use
Perelomov's general definition of group-related coherent states
\cite{Per1}, where it is assumed that the label space $\mathcal{X}$
has a group structure:

\textit{Let $\mathcal{H}$ be a separable Hilbert space, $G$ be a
group, $T(g)$ be an arbitrary representation of group $G$ on the
Hilbert space $\mathcal{H}$, and $|0\rangle$ be an arbitrary
normalized vector in $\mathcal{H}$. Then the set of states $|g
\rangle$ defined by
\begin{equation}\label{defdef}
  |g\rangle = T(g)|0\rangle, \quad  g \in G,
  \end{equation}
is called the system of coherent states related to representation
$T$. The state $|0\rangle$ is called the vacuum state.}

Our families of coherent states will be of Perelomov type, generated
by projective representations of the group $\mathbb{Z} \times U(1)$
and its universal covering with obvious actions on the quantum phase
space $\mathbb{Z} \times S^{1}$. Using basic operators of Borel
quantum kinematics we shall construct families of Weyl operators
which will act on a fiducial vacuum state. Then we shall verify the
resolution unity. Among other properties also the inner products of
coherent states will be investigated because we shall try to show
that their overlap never vanishes like for canonical coherent
states. For close comparison with canonical coherent states also the
Heisenberg uncertainty product for our coherent states will be
examined.

In section 2 we define quantum position and momentum observables on
$S^1$. Section 3 is devoted to the construction of coherent states
on the circle. Here using the basic operators of Borel quantum
kinematics a family of Weyl operators is constructed, acting on a
vacuum vector. We follow the method of our paper \cite{TCh1}, where
coherent states on $\mathbb{Z}_{M} \times \mathbb{Z}_{M}$ were
studied. In section 4 some properties of the coherent states are
studied. In the first place the resolution of unity is proved. Then
in section 5 quantizations on $S^{1}$ corresponding to the
Aharonov-Bohm type quantum description are formulated. In this
framework the coherent states are constructed in the same way as in
section 3 and their properties investigated. They satisfy the
resolution of unity, too.

\section{Quantization on the circle}

Let the configuration space be the unit circle $S^1$. According to
\cite{DST01} the Hilbert space of quantum mechanics on $S^{1}$ is
$\mathcal{H}=L^{2}(S^{1},d\varphi)$ where $\varphi$ is the angle.
The position observables are periodic functions of $\varphi$.
 %If the position coordinate is defined as the angle $\varphi$, then
%$\varphi$ has to take real values modulo $2 \pi$. This means that
%the corresponding position observable is a periodic saw-shaped
%function $\varphi \pmod{2\pi}$ on $\mathbb{R}$.
 The momentum operator is defined using the unitary representation
$\hat{V}(\alpha)$ of the group of rotations $U(1)$ of the unit
circle,
\begin{equation}\label{VVV}
  [\hat{V}(\alpha)\psi](\beta) = \psi(\beta-\alpha), \quad \psi \in
  L^{2}(S^{1},d\varphi), \quad \alpha,\beta \in \Real,
\end{equation}
shifting the argument of periodic functions in
$L^{2}(S^{1},d\varphi)$. The momentum operator is then by Stone's
theorem
\begin{equation}\label{}
 \hat{V}(\alpha)=e^{-i\alpha \hat{P}} \quad
\Rightarrow \quad \hat{P} = -i\frac{d}{d\varphi}.
\end{equation}
The position operator
\begin{equation}\label{QQQ}
  [\hat{Q}\psi](\varphi) = \varphi\psi(\varphi),
\end{equation}
formally satisfies usual commutation relation
\begin{equation}\label{}
  [\hat{Q},\hat{P}]= i\hat{I},
\end{equation}
but it is not well-defined on $\mathcal{H}$. As clearly demonstrated
in \cite{KR96}, a satisfactory position observable is not $\hat{Q}$
but the unitary operator $e^{i\hat{Q}}$ used in the next section for
the construction of coherent states.

In the Dirac notation the position operator satisfies
$$
\hat{Q}  = \int _{-\pi}^{\pi}  \varphi  \mket{\varphi}
\mbra{\varphi} d\varphi, \quad \mbox{with} \quad
\sks{\varphi}{\varphi'}= \delta(\varphi-\varphi').
 $$
The  position operator  $\hat{Q}$  has continuous spectrum $\varphi
\in [-\pi, \pi) $  with the corresponding eigenvectors
$\{\mket{\varphi}\}.$  We take the symmetric interval in order that
the vacuum state (\ref{vacuum}) be symmetric around zero. An
arbitrary quantum state \ket{\psi} can be expressed in the form
$$
 \mket{\psi} =  \int _{-\pi}^{\pi} \psi (\varphi) \mket{\varphi}d\varphi,
 \quad \mbox{where} \quad \psi (\varphi)=\sks{\varphi}{\psi}.$$
It is useful to expand the periodic wave function $\psi (\varphi)$
in the Fourier series
\begin{equation}\label{FT-1}
 \psi(\varphi) = \sum_{n \in \mathbb{Z}} a_n e^{in\varphi}
 \end{equation}
with the expansion coefficients
\begin{equation} \label{FT}
a_n = \frac{1}{2\pi}\int_{S^{1}}e^{-in\varphi}\psi(\varphi)d\varphi.
\end{equation}

\section{Construction of coherent states}

In order to define coherent states directly by Perelomov's method
\cite{Per1}, we should first construct a family of Weyl operators
labeled by elements of the group $\mathbb{Z} \times U(1)$. Second,
it is necessary to determine the vacuum vector $|0,0\rangle$. The
Weyl system is here defined similarly as in \cite{TCh1},
\begin{equation}\label{}
  \widehat{W}(m,\alpha) = e^{im\hat{Q}}e^{-i\alpha
  \hat{P}} = e^{im\hat{Q}}\hat{V}(\alpha), \quad \alpha
  \in [-\pi,\pi), \; m \in \mathbb{Z}.
\end{equation}
The factors do not commute
\begin{equation}\label{zyx}
 e^{im\hat{Q}}e^{-i\alpha\hat{P}}=
 e^{im\alpha}e^{-i\alpha\hat{P}}e^{im\hat{Q}},
\end{equation}
but the operator $e^{im\hat{Q}}$ is now well defined on
$\mathcal{H}$,
\begin{equation}\label{expimq}
  e^{im\hat{Q}}\psi(\varphi) = e^{im\varphi}\psi(\varphi).
\end{equation}
Due to (\ref{zyx}), the unitary Weyl operators
$\widehat{W}(m,\alpha)$ form a projective representation of the
group $\mathbb{Z} \times U(1)$.

The vacuum vector $|0,0\rangle$ will be determined in analogy with
canonical coherent states on $L^{2}(\mathbb{R})$. The requirement
that the vacuum state be an eigenvector of the annihilation operator
with eigenvalue $0$ is transcribed like in \cite{TCh1} in
exponential form
\begin{equation}\label{qaip}
   e^{\hat{Q} + i\hat{P}}|0,0\rangle = |0,0\rangle.
\end{equation}
Using the Baker-Campbell-Hausdorff formula  the operator $e^{\hat{Q}
+ i\hat{P}}$ can be separated in the product of operators
$e^{\hat{Q}}$ and $e^{i\hat{P}}$. Condition (\ref{qaip}) leads to a
rather fat Gaussian
\begin{equation}\label{vacuum}
  \langle \varphi |0,0\rangle = \mathcal{A}e^{-\frac{\varphi^{2}}{2}},
 \quad \varphi \in [-\pi , \pi),
\end{equation}
sitting on the `origin' of the circle. Hence the vacuum state is an
element of our Hilbert space, $|0,0\rangle \in
L^{2}(S^{1},d\varphi)$. At $\varphi=\pm \pi$ it is continuous but
its derivative has a small discontinuity ($\approx e^{-5}$). The
normalization constant $\mathcal{A}$ is given by
\begin{equation}\label{AaA}
  \mathcal{A}=\frac{1}{\sqrt{\int_{-\pi}^{\pi}exp(-\varphi^{2})d\varphi}}
  \doteq 0.751128.
\end{equation}

The family of coherent states in $L^{2}(S^{1},d\varphi)$ is now
generated by the action of the system of unitary Weyl operators
$\widehat{W}(m,\alpha)$ on the vacuum state $|0,0\rangle$:
\begin{equation}\label{CSket}
  |m,\alpha \rangle := \widehat{W}(m,\alpha) |0,0 \rangle.
\end{equation}
The functional form of our coherent states is given by
\begin{equation}\label{CS}
  \langle \varphi |m,\alpha \rangle =
  \mathcal{A}e^{im\varphi}e^{-\frac{(\varphi-\alpha)^{2}}{2}},
  \quad \varphi \in [- \pi,\pi ),
\end{equation}
i.e. for $\alpha \neq 0$ they are displaced and phased versions of
(\ref{vacuum}) with a discontinuity at $\varphi=\pm \pi$.

\section{Properties of our coherent states in
           $L^{2}(S^{1},d\varphi)$}

In this section we shall examine several properties of our coherent
states which are known to hold for canonical coherent states on
$L^{2}(\mathbb{R})$ \cite{AAG,Kla,Per1}. First we shall prove

\noindent \textbf{Theorem.} \textit{For coherent states
(\ref{CSket}) the resolution of unity
\begin{equation}\label{rou}
 \sum_{k \in \mathbb{Z}} \int_{S^{1}}|k,\alpha \rangle
  \langle k,\alpha |d\alpha = c\hat{I},
\end{equation}
holds, where $c=2\pi$.}

\textit{Proof.} Let us choose an arbitrary normalized vector $\eta
\in L^{2}(S^{1},d\varphi)$. Then the inner product of $|\eta\rangle$
with some coherent state $|k,\alpha\rangle$ can be written in the
following integral form:
\begin{equation}\label{fouc}
   \langle k,\alpha | \eta \rangle =
\mathcal{A}\int_{S^{1}}e^{-ik\varphi}e^{-\frac{(\varphi -
\alpha)^{2}}{2}}\eta(\varphi)d\varphi.
 \end{equation}
If we denote the operator on the left-hand side of (\ref{rou}) by
$\hat{O}$, then we have
\begin{equation}\label{219}
   [\hat{O}\eta](\omega) = \mathcal{A}^{2}\sum_{k \in \mathbb{Z}}
  \int_{\mathbf{S}^{1}}e^{ik\omega}e^{-\frac{(\omega -\alpha)^{2}}{2}}
[\int_{\mathbf{S}^{1}}e^{-ik\varphi}
e^{-\frac{(\varphi-\alpha)^{2}}{2}}\eta(\varphi)d\varphi]d\alpha.
\end{equation}
Now the expression in the square brackets is in fact $2\pi$ times
the $k$-th expansion coefficient (\ref{FT}) of the Fourier
decomposition of the function $\exp(-\frac{(\varphi -
\alpha)^{2}}{2})\eta(\varphi)$. Applying (\ref{FT-1}) we obtain
\begin{equation}\label{kols}
  [\hat{O}\eta](\omega) = 2\pi \mathcal{A}^{2}
  \int_{\mathbf{S}^{1}}exp(-(\omega -\alpha)^{2})\eta(\omega)d\alpha=
  2\pi\eta(\omega),
\end{equation}
since the integral in (\ref{kols}) yields the squared norm of the
coherent state $|m,\alpha\rangle$. $\Box$

Secondly, inner products (overlaps) of our normalized coherent
states will be studied. Here it is necessary to correctly realize
the way how the operator $\hat{V}(\alpha) = \exp(-i\alpha\hat{P})$
acts on the Hilbert space $L^{2}(S^{1},d\varphi)$ when the circle
$S^{1}$
--- the configuration space --- is identified with the interval
$[-\pi,\pi)$. Then the action of operator $e^{-i\alpha\hat{P}}$ on
function $\psi(\varphi) \in L^{2}(S^{1},d\varphi)$  for
 $\alpha \in [0,\pi)$ has the form:
\begin{eqnarray}\label{pro1}
  e^{-i\alpha\hat{P}}\psi(\varphi) & = &
      \psi(\varphi - \alpha) \quad \mbox{for} \quad
      \varphi \in [-\pi + \alpha,\pi), \\
  e^{-i\alpha\hat{P}}\psi(\varphi) & = &
    \psi(\varphi - \alpha +2\pi) \quad \mbox{for} \quad
    \varphi \in [-\pi , -\pi + \alpha).
 \end{eqnarray}
For $\alpha \in [-\pi,0)$ we have
\begin{eqnarray}\label{pro2}
  e^{-i\alpha\hat{P}}\psi(\varphi) & = &
     \psi(\varphi - \alpha) \quad \mbox{for} \quad \varphi \in [-\pi ,\pi + \alpha), \\
 e^{-i\alpha\hat{P}}\psi(\varphi) & = &
    \psi(\varphi - \alpha -2\pi) \quad \mbox{for} \quad \varphi \in [\pi + \alpha , \pi).
 \end{eqnarray}
Note that one has to consider addition modulo $2\pi$ in the argument
of function $\psi$. For this reason inner products cannot be
calculated simply according to
\begin{equation}\label{}
  \langle m,\alpha| n,\beta \rangle =
\mathcal{A}^{2}\int_{-\pi}^{\pi}e^{-i\varphi(n-m)}
e^{-\frac{(\varphi-\alpha)^{2}}{2}}e^{-\frac{(\varphi -
  \beta)^{2}}{2}}d\varphi.
\end{equation}

From now on we shall restrict ourselves only to the cases when
$\alpha$ and $\beta$ are non-negative numbers:
\begin{equation}\label{230}
  \alpha \in [0, \pi), \quad \beta \in [0,\pi).
\end{equation}
Without loss of generality we may also assume
\begin{equation}\label{231}
  \beta \geq \alpha.
\end{equation}
Taking into account (\ref{pro1}) --- (\ref{pro2}), we split the
inner product of two coherent states into two terms
\begin{equation}\label{232}
 \langle m,\alpha| n,\beta\rangle =
 \mathcal{A}^{2}I_{1}(\alpha,\beta,n-m) +
  \mathcal{A}^{2}I_{2}(\alpha,\beta,n-m),
\end{equation}
where
\begin{equation}\label{I1}
  I_{1}(\alpha,\beta,n-m) := \int_{\alpha - \pi}^{\beta -
  \pi}e^{i\varphi(n-m)} e^{-\frac{(\varphi - \alpha)^{2}}{2}}
  e^{-\frac{(\varphi-\beta + 2\pi)^{2}}{2}}d\varphi
\end{equation}
and
\begin{equation}\label{}
   I_{2}(\alpha,\beta,n-m) := \int_{\beta - \pi}^{\pi +
  \alpha}e^{i\varphi(n-m)} e^{-\frac{(\varphi-\alpha)^{2}}{2}}
   e^{-\frac{(\varphi-\beta)^{2}}{2}} d\varphi.
\end{equation}
The integrals $I_{1}(\alpha,\beta,n-m)$ and
$I_{2}(\alpha,\beta,n-m)$ can be expressed in terms of the error
function of a complex variable $z$,
\begin{equation}\label{erf}
  \mbox{erf}(z) := \frac{2}{\sqrt{\pi}}\int_{\Gamma
  (z)}e^{-\psi^{2}}d\psi;
\end{equation}
here $\Gamma(z)$ denotes an arbitrary continuous path of finite
length which connects the origin $0 \in \mathbb{C}$ with complex
number $z \in \mathbb{C}$. Since the Gauss function is analytic, the
definition (\ref{erf}) does not depend on the path $\Gamma (z)$.

The integral $I_{1}(\alpha,\beta,n-m)$, after the substitution
\begin{equation}\label{}
  \omega = \varphi + \pi - \frac{\alpha + \beta}{2},
\end{equation}
takes the form
\begin{eqnarray}\label{}
  \nonumber I_{1}(\alpha,\beta,n-m)
 &= &
 e^{-(\frac{\beta-\alpha}{2})^{2}-\pi}e^{(\frac{\alpha+\beta}{2}-\pi)(m-n)}
 \times \\ & & \times
 \int_{\frac{\alpha-\beta}{2}}^{\frac{\beta-\alpha}{2}}e^{i\omega(n-m)}
 e^{-\omega^{2}}d\omega,
\end{eqnarray}
which leads to the formula
\begin{eqnarray}\label{II11}
  \nonumber I_{1}(\alpha,\beta,n-m) & = &
(-\frac{\sqrt{\pi}}{2})e^{-(\frac{\beta-\alpha}{2})^{2}-\pi}
e^{i(\frac{\alpha+\beta}{2}-\pi)(m-n)}e^{-\frac{(n-m)^{2}}{4}}\times
\\ & &
\times[\mbox{erf}(\frac{\alpha-\beta}{2}+\frac{i(n-m)}{2})+\\
& &+ \mbox{erf}(\frac{\alpha-\beta}{2}-\frac{i(n-m)}{2})].\nonumber
\end{eqnarray}
The other integral $I_{2}(\alpha,\beta,n-m)$, after the substitution
$  \omega = \varphi-\frac{\alpha+\beta}{2}$, yields
\begin{eqnarray}\label{II22}
  \nonumber I_{2}(\alpha,\beta,n-m) & = &
(-\frac{\sqrt{\pi}}{2})e^{-(\frac{\beta-\alpha}{2})^{2}}
e^{i(\frac{\alpha+\beta}{2})(m-n))}e^{-\frac{(n-m)^{2}}{4}}\times \\
& & \times[\mbox{erf}(\frac{\alpha-\beta}{2}-\pi+\frac{i(n-m)}{2})+\\
& &
+\mbox{erf}(\frac{\alpha-\beta}{2}-\pi-\frac{i(n-m)}{2})].\nonumber
\end{eqnarray}

We have to admit that, unfortunately, we do not see any way how to
further simplify the above analytic expressions of the integrals
$I_{1}(\alpha,\beta,n-m)$ and $I_{2}(\alpha,\beta,n-m)$ to see
whether the coherent states are mutually non-orthogonal. However, we
have numerically computed absolute values of the inner products for
many pairs of coherent states and have plotted the graphs of the
absolute value of the inner product for several different values of
$n-m$ fixed in each graph. It was apparent --- for all plotted cases
--- that the overlap never vanishes. For convenience of the reader we
attach in figures 1--3 the graphs of the absolute value of the
overlap as function of parameters $\alpha$ and $\beta$ (parameter
$n-m$ is fixed for each graph) confirming this property.

\begin{center}
\begin{figure}
 \rotatebox{-90}{
   \includegraphics[scale=0.5]{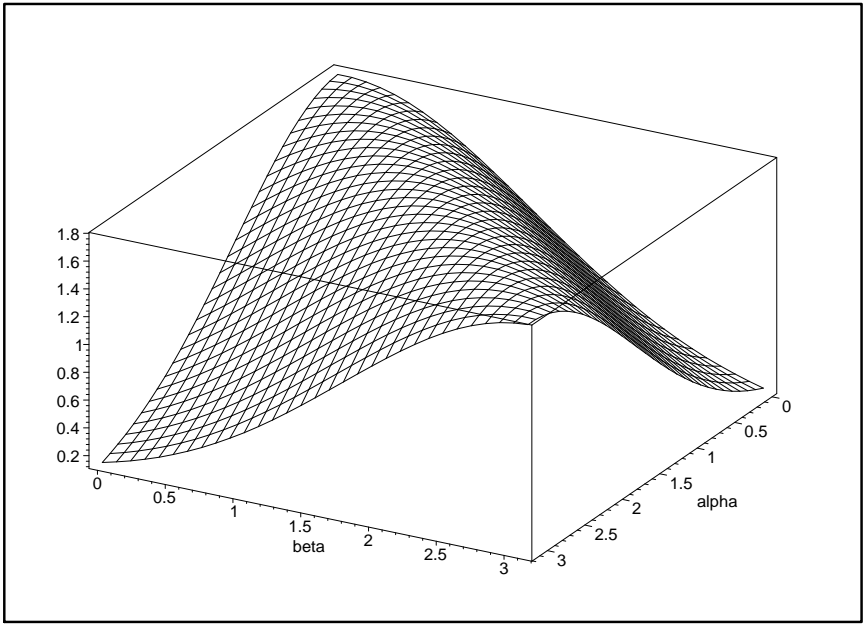}
 }
 \caption{Inner product of coherent states on the circle for $m-n=0$.}
 \end{figure}
 \end{center}
  \begin{figure}
  \rotatebox{-90}{
   \includegraphics[scale=0.5]{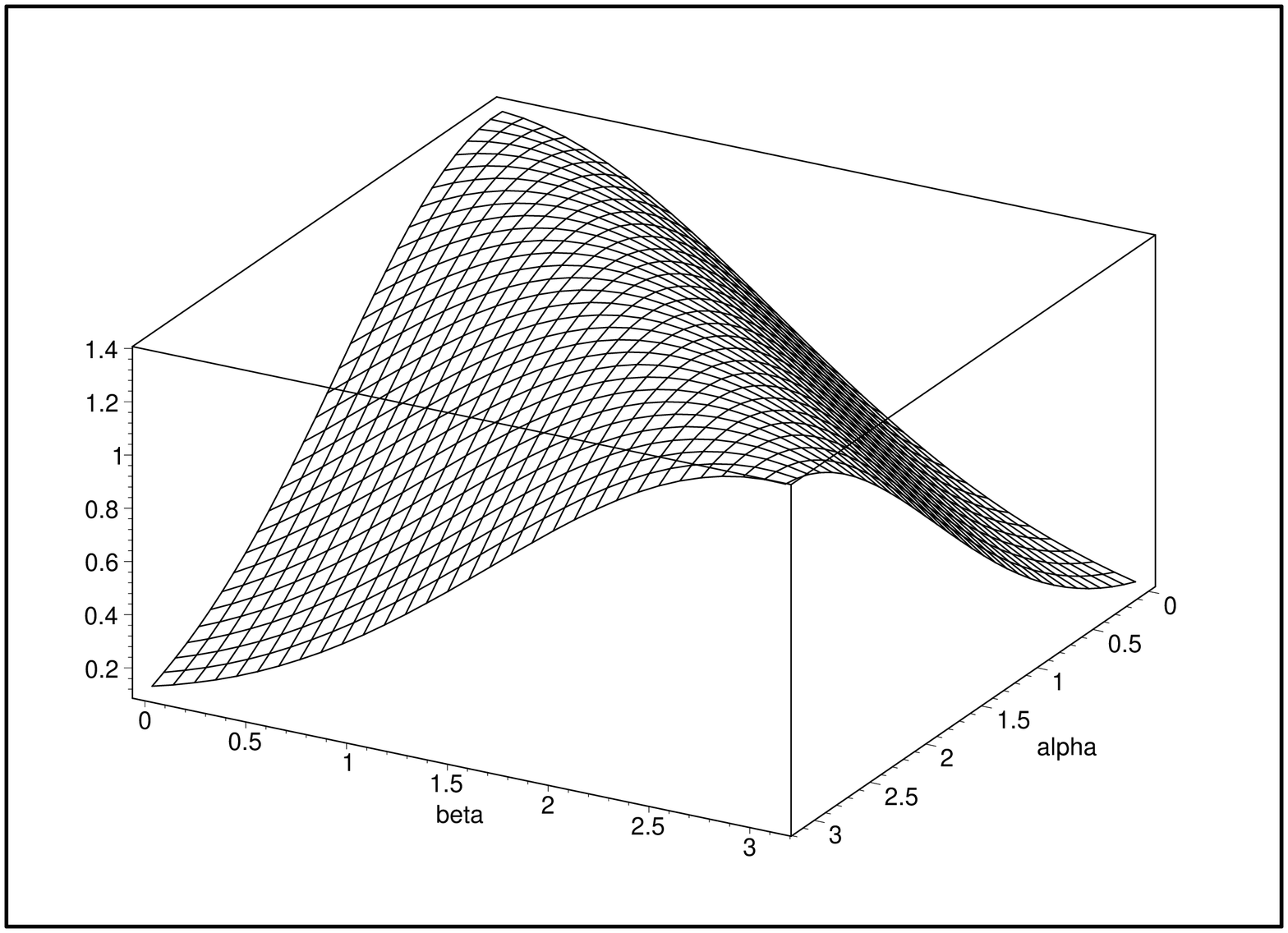}
 }
 \caption{Inner product of coherent states on the circle for $m-n=1$.}
  \end{figure}
 \begin{figure}
  \rotatebox{-90}{
   \includegraphics[scale=0.5]{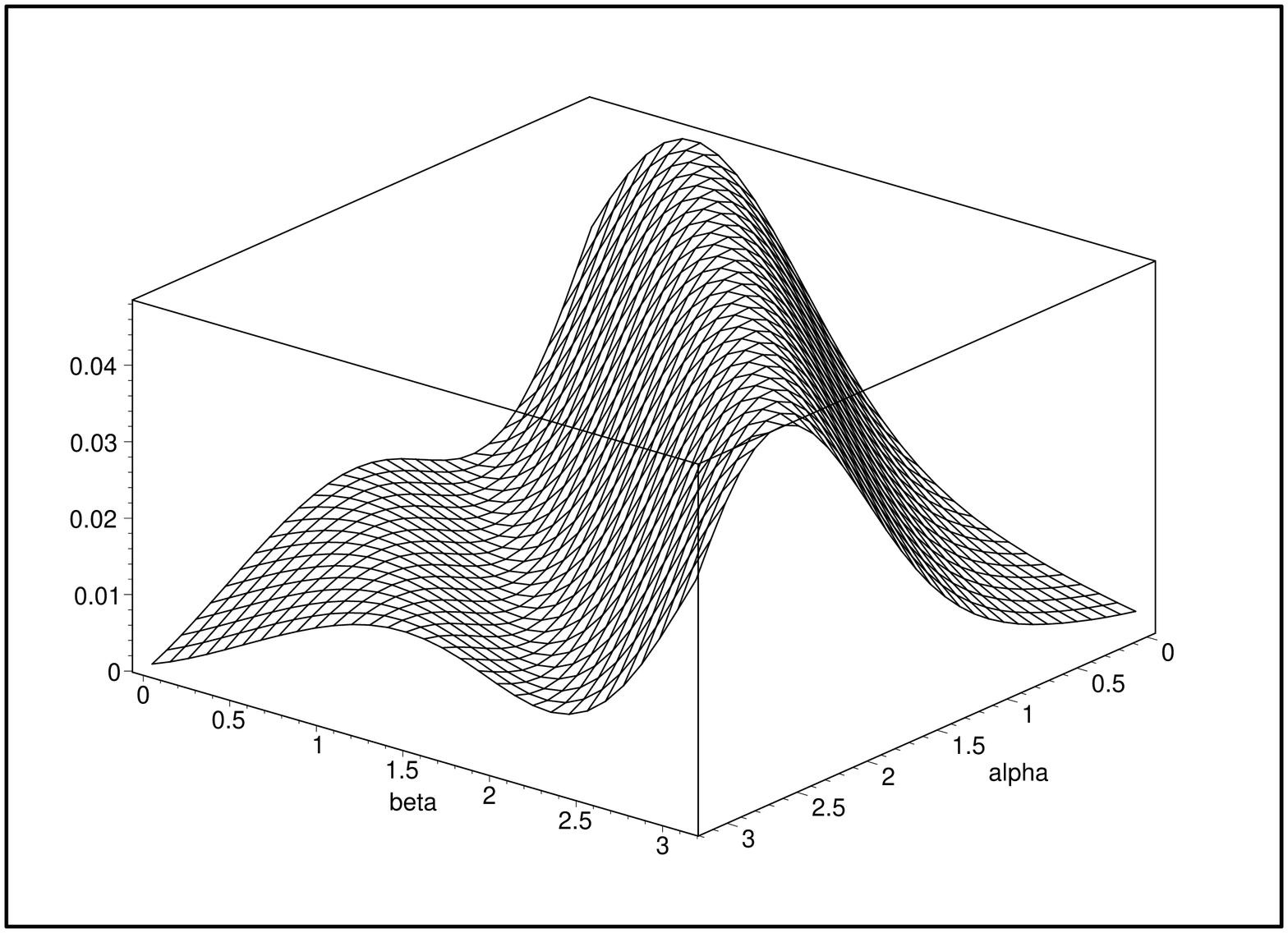}
 }
 \caption{Inner product of coherent states on the circle for $m-n=4$.}
 \end{figure}

The next checked quantities are the expectation values of basic
observables in the coherent states $|m,\alpha\rangle$ for $\alpha
\geq 0$. Explicitly we have for the position operator
\begin{equation}\label{244}
  \langle m,\alpha|\hat{Q}|m,\alpha\rangle = \mathcal{A}^{2}\int_{-\pi}^{-\pi +
  \alpha}\varphi e^{-(\varphi-\alpha + 2\pi)^{2}}d\varphi + \mathcal{A}^{2}\int_{-\pi +
  \alpha}^{\pi}\varphi e^{-(\varphi -\alpha)^{2}}d\varphi,
\end{equation}
or, using the error function,
\begin{equation}\label{245}
  \langle m,\alpha|\hat{Q}|m,\alpha\rangle = \alpha +
  q_{1}^{(+)}(\alpha),
\end{equation}
where
\begin{equation}\label{2451}
 q_{1}^{(+)}(\alpha)= \mathcal{A}^{2}
\sqrt{\pi^{3}}(\mbox{erf}(\pi)-\mbox{erf}(\pi - \alpha)).
\end{equation}
One observes that the expectation values do not depend on $m$ (this
result is obtained also for $\alpha$ negative). Note that the
expectation value of position is nonlinear in $\alpha$ for $\alpha
\neq 0$. It is clear that for $\alpha =0$ the quasi-Gaussian
(\ref{vacuum}) is symmetric around $\varphi = 0$, so the expectation
value is an integral of an odd function, evidently equal to zero.
However, if $\alpha \neq 0$, the displaced quasi-Gaussian is no more
symmetric around $\varphi = 0$ and the integration leads to a
deviation depending on a difference of error functions (\ref{2451});
maximal deviation is attained for $\alpha \rightarrow \pi$.

\noindent \footnotesize \textbf{Remark} \cite{R}. If instead of the
expectation values of $\hat{Q}$ expectation values of the unitary
position operator $e^{i\hat{Q}}$ are considered in our coherent
states, we obtain using (\ref{zyx})
\begin{equation}\label{}
  \langle m,\alpha|e^{i\hat{Q}}|m,\alpha\rangle =
  e^{i\alpha} \langle 0,0|e^{i\hat{Q}}|0,0\rangle .
\end{equation}
Thus the relative expectation value \cite{KR96}
\begin{equation}\label{}
  \frac{\langle m,\alpha|e^{i\hat{Q}}|m,\alpha\rangle}{
   \langle 0,0|e^{i\hat{Q}}|0,0\rangle} = e^{i\alpha},
\end{equation}
yields the exact value of the classical angle, but
\begin{equation}
\langle 0,0|e^{i\hat{Q}}|0,0\rangle \doteq 0.778816
\end{equation}
does not lie on the unit circle. Taking into account that $e^{-1/4}
\doteq 0.778816$, it turns out that our coherent states yield a very
good approximation to formula (3.52) of \cite{KR96}.

\normalsize The expectation value of the square of position operator
in state $|m,\alpha\rangle$ is
\begin{equation}\label{248}
\langle m,\alpha|\hat{Q}^{2}|m,\alpha\rangle =
\mathcal{A}^{2}\int_{-\pi}^{-\pi + \alpha}\varphi^{2}
e^{-(\varphi-\alpha + 2\pi)^{2}}d\varphi + \mathcal{A}^{2}\int_{-\pi
+
 \alpha}^{\pi}\varphi^{2} e^{-(\varphi -\alpha)^{2}}d\varphi.
\end{equation}
The computation gives
\begin{equation}\label{}
  \langle m,\alpha|\hat{Q}^{2}|m,\alpha\rangle =
 \alpha^{2}+\frac{1}{2} + q_{2}^{(+)}(\alpha),
\end{equation}
where
\begin{equation}\label{250}
q_{2}^{(+)}(\alpha) =
\mathcal{A}^{2}[\pi(e^{-\pi^{2}}-2e^{-(\pi-\alpha)^{2}})+2\sqrt{\pi^{3}}(\pi
- \alpha)(\mbox{erf}(\pi)-\mbox{erf}(\pi-\alpha))].
\end{equation}

Important among the checked quantities are also the expectation
values of the momentum operator in coherent states
$|m,\alpha\rangle$. The explicit form
\begin{eqnarray}\label{}
  \nonumber \langle m,\alpha|\hat{P}|m,\alpha\rangle &=& \mathcal{A}^{2}\int_{-\pi}^{-\pi +
  \alpha}(m +i(\varphi - \alpha + 2\pi)) e^{-(\varphi-\alpha + 2\pi)^{2}}d\varphi
  \\ \nonumber
   &+& \mathcal{A}^{2}\int_{-\pi +
  \alpha}^{\pi}(m+i(\varphi-\alpha)) e^{-(\varphi
  -\alpha)^{2}}d\varphi
\end{eqnarray}
can be simplified into the formula
\begin{equation}\label{}
  \langle m,\alpha|\hat{P}|m,\alpha\rangle = m,
\end{equation}
which could be anticipated by analogy with the canonical coherent
states on $L^{2}(\mathbb{R})$.

Finally, the expectation values of the square of momentum operator
in coherent states $|m,\alpha\rangle$ are determined by computing
the integrals
\begin{eqnarray}\label{}
 \nonumber \langle m,\alpha|\hat{P}^{2}|m,\alpha\rangle &=&
\mathcal{A}^{2}\int_{-\pi}^{-\pi + \alpha}[1+(m +i(\varphi -
\alpha + 2\pi))^{2}] e^{-(\varphi-\alpha + 2\pi)^{2}}d\varphi  \\
&+& \mathcal{A}^{2}\int_{-\pi +
  \alpha}^{\pi}[1+(m+i(\varphi-\alpha))^{2}] e^{-(\varphi
  -\alpha)^{2}}d\varphi.
\end{eqnarray}
The result
\begin{equation}\label{p2}
  \langle m,\alpha|\hat{P}^{2}|m,\alpha\rangle =
 m^{2}+\frac{1}{2} +p_{2}^{(+)}, \quad\mbox{where} \quad
 p_{2}^{(+)}=\mathcal{A}^{2}\pi e^{-\pi^{2}},
\end{equation}
does not depend on $\alpha$.

Now these expectation values can be used to form Heisenberg's
uncertainty product for our coherent states. First, the dispersion
of the position operator is
\begin{eqnarray}\label{}
 \nonumber \Delta_{|m,\alpha\rangle}\hat{Q} &=&
 \sqrt{\langle m,\alpha|\hat{Q}^{2}|m,\alpha\rangle - \langle
 m,\alpha|\hat{Q}|m,\alpha\rangle^{2}}   \\
  &=& \sqrt{\frac{1}{2}+q_{2}^{(+)}(\alpha)-2\alpha
  q_{1}^{(+)}(\alpha)-q_{1}^{(+)}(\alpha)^{2}}.
\end{eqnarray}
This result is independent of $m$, it depends only on $\alpha$. If
both $q_{1}^{(+)}$ and $q_{2}^{(+)}$ were zero, the same value as
for canonical coherent states on $L^{2}(\mathbb{R})$ would result,
namely $\sqrt{\frac{1}{2}}$. However, the functions $q_{1}^{(+)}$
and $q_{2}^{(+)}$ do not vanish except $q_{1}^{(+)}(0)=0$. Second,
the dispersion of the momentum operator in state $|m,\alpha\rangle$
is independent of $m$ and $\alpha$ and different from
$\sqrt{\frac{1}{2}}$,
\begin{equation}\label{}
 \Delta_{|m,\alpha\rangle}\hat{P} =
 \sqrt{\langle m,\alpha|\hat{P}^{2}|m,\alpha\rangle - \langle
 m,\alpha|\hat{P}|m,\alpha\rangle^{2}} =
 \sqrt{\frac{1}{2}+p_{2}^{(+)}}.
\end{equation}
So finally the Heisenberg uncertainty product has the form
\begin{equation}\label{relneup}
   \Delta_{|m,\alpha\rangle}\widehat{Q}\cdot
  \Delta_{|m,\alpha\rangle}\widehat{P} =\sqrt{\frac{1}{2} +p_{2}^{(\pm)}}
 \sqrt{\frac{1}{2}+q_{2}^{(\pm)}(\alpha)-2\alpha
  q_{1}^{(\pm)}(\alpha)-q_{1}^{(\pm)}(\alpha)^{2}},
\end{equation}
where superscripts are included also for $\alpha \leq 0$. The
results for $\alpha \leq 0$ are given below for completeness:
\begin{equation}\label{}\nonumber
   q_{1}^{(-)}(\alpha) = \mathcal{A}^{2}\sqrt{\pi^{3}}
   (\mbox{erf}(\pi)- \mbox{erf}(\pi + \alpha)),
\end{equation}
\begin{equation}\label{}\nonumber
  q_{2}^{(-)}(\alpha) = \mathcal{A}^{2}[\pi(e^{-\pi^{2}}-
  2e^{-(\pi-\alpha)^{2}})+2\sqrt{\pi^{3}}(\pi +
 \alpha)(\mbox{erf}(\pi)-\mbox{erf}(\pi+\alpha))],
\end{equation}
and
\begin{equation}\label{}\nonumber
  p_{2}^{(-)} =\mathcal{A}^{2}\pi \exp(-\pi^{2}) = p_{2}^{(+)}.
\end{equation}

The plot of $\Delta_{|m,\alpha\rangle}\hat{Q}\cdot
\Delta_{|m,\alpha\rangle}\hat{P}$ as function of non-negative
parameter $\alpha$ is given in figure 4. The graph for $\alpha$
negative is obtained as its even prolongation to $\alpha \leq 0$.
\begin{figure}
 \rotatebox{-90}{
   \includegraphics[scale=0.4]{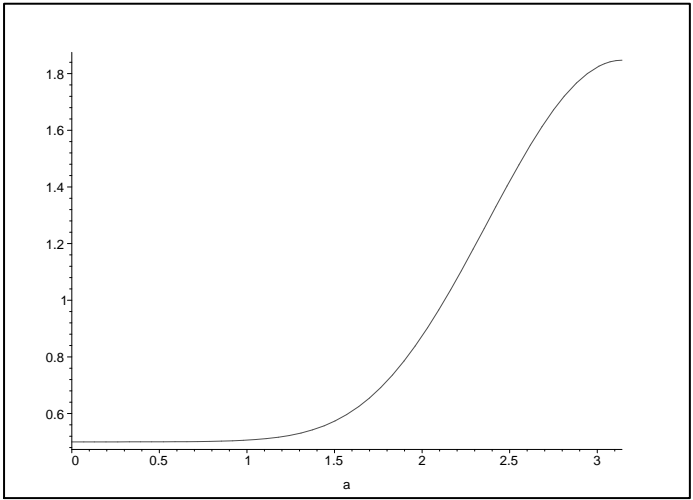}
 }
 \caption{Heisenberg uncertainty product for positive values of $\alpha$.}
 \end{figure}
One can see that $\Delta_{|m,\alpha\rangle}\hat{Q}\cdot
  \Delta_{|m,\alpha\rangle}\hat{P}$ achieves its minimum for
$\alpha = 0$, i.e. with
\begin{equation}\label{}
  q_{1}^{(\pm)}(0) = 0, \; q_{2}^{(\pm)}(0) =
  -\mathcal{A}^{2}\pi e^{-\pi^{2}}, \;
  p_{2}^{(\pm)}=\mathcal{A}^{2}\pi e^{-\pi^{2}}.
\end{equation}
Using (\ref{relneup}) we get
\begin{equation}\label{Heis}
 \nonumber \Delta_{|m,0\rangle}\hat{Q}\cdot
  \Delta_{|m,0\rangle}\hat{P}  =\sqrt{\frac{1}{2} +p_{2}^{(\pm)}}\cdot
  \sqrt{\frac{1}{2}+q_{2}^{(\pm)}(0)} =
   \sqrt{\frac{1}{4} - \mathcal{A}^{2}\pi^{2}
  e^{-2\pi^{2}}} < \frac{1}{2}.
\end{equation}
Numerical evaluation gives for $\alpha =0$ the actual value
$\Delta_{|m,0\rangle}\hat{Q}\cdot
  \Delta_{|m,0\rangle}\hat{P} \doteq 0.4999999973$, i.e. very
slightly below the Heisenberg limit. This circumstance will be
discussed in section 6.

\section{Quantizations of the Aharonov-Bohm type and coherent states}

Let us now describe the quantizations on $S^1$ of the Aharonov-Bohm
type in the way delineated in \cite{DST01}. As mentioned in the
introduction, inequivalent quantizations are labeled by $\theta \in
[0,1)$. They correspond to quantum mechanics of a particle of charge
$e$ confined to a circle through which a magnetic flux tube
penetrates. The relation of the magnetic flux $\Phi$ to parameter
$\theta$ is given by
 $$
    2\pi \theta = \frac{e}{\hbar}\Phi.
 $$
As before, also in the following we set $\hbar=1$.

The simplest way to obtain the set of quantizations labeled by
$\theta$ is to replace the group $U(1)$ by its simply connected
universal covering group $\mathbb{R}$ \cite{DST01}. Action $\sigma$
of $\mathbb{R}$ on $S^{1}$ is natural,
\begin{equation}\label{}
  \sigma:\mathbb{R}\times S^{1}\rightarrow S^{1}: \;
  (x,e^{i\varphi})\mapsto e^{i(x+\varphi)}, \quad
  e^{i\varphi} \in S^{1}, \; x \in \mathbb{R}.
\end{equation}
It is evidently transitive and the stability subgroup is
$\mathbb{Z}$. The set of all inequivalent irreducible unitary
representations of $\mathbb{Z}$ is labeled by parameter $\theta \in
[0,1)$,
\begin{equation}\label{}
  L^{\theta}: \mathbb{Z} \rightarrow U(1): \; n \mapsto
  e^{2\pi in\theta}.
\end{equation}
These one-dimensional representations classify inequivalent quantum
mechanics labeled by parameter $\theta \in [0,1)$. The Hilbert space
$\mathcal{H}^{\theta}$ corresponding to parameter $\theta$ contains
Borel complex functions $\chi(x)$ on $\mathbb{R}$ with finite norm
which are quasi-periodic,
\begin{equation}\label{hsf}
  \chi(x+2n\pi)=L^{\theta}(n)^{-1}\chi(x)=e^{-2\pi i n\theta}\chi(x).
\end{equation}
The inner product in $\mathcal{H}^{\theta}$ is
\begin{equation}\label{}
  (\psi,\chi)_\theta = \int_{a}^{a+2\pi}\overline{\psi(x)}\chi(x)dx, \quad
  a \in \mathbb{R}, \quad \psi,\chi \in \mathcal{H}^{\theta}.
\end{equation}
The induced unitary representation of the group $\mathbb{R}$ on
$\mathcal{H}^{\theta}$ has the simple form
\begin{equation}\label{}
[\hat{V}^{\theta}(\alpha)\chi](\beta) =
e^{-i\alpha\hat{P}^{\theta}}\chi(\beta)= \chi (\beta - \alpha),
\quad \alpha\in  \mathbb{R},
\end{equation}
hence the momentum operator $\hat{P}^{\theta}$ is simply
\begin{equation}\label{}
  \hat{P}^{\theta} = -i\frac{d}{dx}.
\end{equation}
The position operator on $\mathcal{H}^{\theta}$ is more complicated.
It is the multiplication by a saw-shaped function on $\mathbb{R}$,
\begin{equation}\label{}
  (\hat{Q}\chi)(x)=(x \pmod{2\pi})\chi(x), \quad \chi \in
  \mathcal{H}^{\theta},
\end{equation}
so that function $\hat{Q}\chi(x)$ remains quasi-periodic
(\ref{hsf}). The momentum operators $\widehat{P}^{\theta}$ have the
same form for all $\theta$.

For our calculations it is advantageous to identify the Hilbert
spaces $\mathcal{H}^{\theta}$ with the Hilbert space
$L^{2}(S^{1},d\varphi)$ of periodic functions $\psi(\varphi)$ via a
gauge transformation
\begin{equation}\label{}
  \mathcal{U}^{\theta}: \mathcal{H}^{\theta} \rightarrow
  L^{2}(S^{1},d\varphi): \;  \chi(\varphi) \mapsto\psi(\varphi)=
  e^{i\theta \varphi} \chi(\varphi).
\end{equation}
The operator $\hat{P}^{\theta}$ on $\mathcal{H}^{\theta}$ is
transformed to
\begin{equation}\label{momentumtheta}
  \widehat{\mathbf{P^{\theta}}}=
  \mathcal{U}^{\theta}\hat{P}^{\theta}(\mathcal{U}^{\theta})^{-1}
  = -i\frac{d}{d\varphi} - \theta.
\end{equation}
This covariant derivative on $L^{2}(S^{1},d\varphi)$ includes a
constant vector potential $A= \Phi / 2\pi$ corresponding to the
Aharonov-Bohm magnetic flux $\Phi = \int_{S^1}A d\varphi$. The
position operator acts by multiplying by independent variable
(\ref{QQQ}).

Let us now define families of coherent states for quantum mechanics
labeled by $\theta$. We shall proceed as in the previous chapter
working in the Hilbert space $L^{2}(S^{1},d\varphi)$ of periodic
functions. It is easy to see that the commutation relation
(\ref{zyx}) holds in the same form
\begin{equation}\label{}
e^{im\hat{Q}}e^{-i\alpha\widehat{\mathbf{P^{\theta}}}}=
e^{im\alpha}e^{-i\alpha\widehat{\mathbf{P^{\theta}}}}e^{im\hat{Q}},
\quad \alpha \in [-\pi,\pi),  \quad m \in \mathbb{Z}.
\end{equation}
For the vacuum vector we solve
\begin{equation}\label{}
   e^{\hat{Q} + i\widehat{\mathbf{P^{\theta}}}}|0,0,\theta \rangle =
  |0,0,\theta \rangle
\end{equation}
and find the vacuum state
\begin{equation}\label{}
  \langle \varphi|0,0,\theta \rangle =
  \mathcal{A}_{\theta}e^{-(\varphi-i\theta)^{2}/2},
\end{equation}
where the normalization constant
$$\mathcal{A}_{\theta}=\mathcal{A}e^{-\theta^{2} /2}$$
(for $\mathcal{A}$ see (\ref{AaA})). The coherent states are now
defined by the action of Weyl operators
\begin{equation}\label{doop}
  \widehat{W}^{\theta}(m,\alpha)=
  e^{im\hat{Q}}e^{-i\alpha\widehat{\mathbf{P^{\theta}}}}
\end{equation}
 on the vacuum vector and have the following explicit functional form:
\begin{eqnarray}\label{CStheta}
 \nonumber  \langle \varphi|m,\alpha,\phi\rangle & = &
 \langle \varphi \vert \widehat{W}^{\theta}(m,\alpha)|m,\alpha,\theta\rangle = \\
 & = &
\mathcal{A}_{\theta}e^{im\varphi}e^{-(\varphi-\alpha-i\theta)^{2}/2},
\quad \varphi \in [-\pi,\pi).
\end{eqnarray}

Now concerning the properties of coherent states for quantum
mechanics labeled by parameter $\theta$, we start with

\noindent \textbf{Theorem.} \textit{For coherent states
(\ref{CStheta}) the resolution of unity}
\begin{equation}\label{routheta}
 \sum_{k \in \mathbb{Z}} \int_{S^{1}}|k,\alpha,\theta \rangle
  \langle k,\alpha,\theta |d\alpha = c\hat{I},
\end{equation}
\textit{holds with $c=2\pi$.}

\textit{Proof.} For the proof we take the operator
$\hat{O}^{\theta}$ on the left-hand side of (\ref{routheta}) and let
it act on arbitrary normalized function $\eta \in
L^2(S^{1},d\varphi)$
\begin{eqnarray}\label{}\nonumber
 [\hat{O}^{\theta}\eta](\omega) = \\
\mathcal{A}_{\theta}^{2}\sum_{k \in
\mathbb{Z}} \int_{S^{1}}e^{ik\omega}e^{-(\omega -\alpha -
i\theta)^{2} /2} \cdot
[\int_{S^{1}}e^{-ik\varphi}e^{-(\varphi-\alpha + i\theta)^{2}
/2}\eta(\varphi)d\varphi]d\alpha.
\end{eqnarray}
If we perform similar computation as in (\ref{219}), we finally
obtain
\begin{equation}\label{}
[\hat{O}^{\theta}\eta](\omega) = 2\pi \mathcal{A}_{\theta}^{2}
\eta(\omega)  \int_{S^{1}}e^{-(\omega - \alpha -i\theta)^{2}}d\alpha
=  2\pi \eta(\omega).  \Box
\end{equation}

Next, we briefly examine the overlaps of the coherent states. We
will keep the restrictions (\ref{230}), (\ref{231}) on parameters
$\alpha$ and $\beta$, and then divide the inner product in two
integrals. Proceeding as in (\ref{232})
\begin{equation}\label{}
\langle m,\alpha,\theta| n,\beta,\theta\rangle =
\mathcal{A}_{\theta}^{2}I_{1}(\alpha,\beta,n-m,\theta) +
  \mathcal{A}_{\theta}^{2}I_{2}(\alpha,\beta,m-n,\theta),
\end{equation}
we have
\begin{equation}\label{}
  I_{1}(\alpha,\beta,n-m,\theta) = \int_{\alpha - \pi}^{\beta -
  \pi}e^{i\varphi(n-m)}e^{-\frac{(\varphi -
  \alpha + i\theta)^{2}}{2}}e^{-\frac{(\varphi-\beta +
  2\pi - i\theta)^{2}}{2}}d\varphi
\end{equation}
and
\begin{equation}\label{}
   I_{2}(\alpha,\beta,n-m,\theta) = \int_{\beta - \pi}^{\pi +
\alpha}e^{i\varphi(n-m)}e^{-\frac{(\varphi-\alpha +
i\theta)^{2}}{2}}e^{-\frac{(\varphi-\beta -
i\theta)^{2}}{2}}d\varphi.
\end{equation}
Computation of $ I_{1}(\alpha,\beta,n-m,\theta)$ and
$I_{2}(\alpha,\beta,n-m,\theta)$ gives us
\begin{eqnarray}\label{}
\nonumber  I_{1}(\alpha,\beta,n-m,\theta) = \\
 e^{\theta^{2}}e^{-i\theta(\beta - \alpha +2\pi)}
 (-\frac{\sqrt{\pi}}{2})e^{-(\frac{\beta-\alpha}{2})^{2}-\pi}
 e^{i(\frac{\alpha+\beta}{2}-\pi)(m-n)}e^{-\frac{(n-m)^{2}}{4}}\times
\nonumber \\ \times
[\mbox{erf}(\frac{\alpha-\beta}{2}+\frac{i(n-m)}{2})+
\mbox{erf}(\frac{\alpha-\beta}{2}-\frac{i(n-m)}{2})]
\end{eqnarray}
and
\begin{eqnarray}\label{}
\nonumber  I_{2}(\alpha,\beta,n-m,\theta) =\\
 e^{\theta^{2}}e^{-i\theta(\beta -
 \alpha)}(-\frac{\sqrt{\pi}}{2})e^{-(\frac{\beta-\alpha}{2})^{2}}
 e^{i(\frac{\alpha+\beta}{2})(m-n))}e^{-\frac{(n-m)^{2}}{4}}\times
 \nonumber  \\ \times [\mbox{erf}(\frac{\alpha-\beta}{2}-\pi+\frac{i(n-m)}{2})+
\mbox{erf}(\frac{\alpha-\beta}{2}-\pi-\frac{i(n-m)}{2})].
\end{eqnarray}
Comparing these results with (\ref{II11}) and (\ref{II22}), we can
write
\begin{equation}\label{}
  I_{1}(\alpha,\beta,n-m,\theta) =
 e^{\theta^{2}}e^{-i\theta(\beta - \alpha +2\pi)}I_{1}(\alpha,\beta,n-m)
\end{equation}
and
\begin{equation}\label{}
  I_{2}(\alpha,\beta,n-m,\theta) =
  e^{\theta^{2}}e^{-i\theta(\beta -\alpha)}I_{2}(\alpha,\beta,n-m).
\end{equation}
The inner product for two coherent states is finally
\begin{eqnarray}\label{}
   \nonumber \langle m,\alpha,\theta| n,\beta,\theta\rangle = \\
  =\mathcal{A}^{2} e^{-i\theta(\beta - \alpha +2\pi)}I_{1}(\alpha,\beta,n-m) +
   \mathcal{A}^{2} e^{-i\theta(\beta -\alpha)}I_{2}(\alpha,\beta,m-n).
\end{eqnarray}

The expectation values of position and momentum operators and their
squares were also computed. For $\alpha \geq 0$ the correction
functions in (\ref{2451}), (\ref{250}) and (\ref{p2}) are changed to
\begin{eqnarray}\label{}
 \nonumber q_{1}^{(+)\theta}(\alpha)& =&
 \mathcal{A}_{\theta}^{2}\sqrt{\pi^{3}}(\mbox{erf}(\pi)-\mbox{erf}(\pi -\alpha)),
 \\ \nonumber q_{2}^{(+)\theta}(\alpha) &=&
 \mathcal{A}_{\theta}^{2}[\pi(e^{-\pi^{2}}-2e^{-(\pi-\alpha)^{2}})+
 2\sqrt{\pi^{3}}(\pi - \alpha)(\mbox{erf}(\pi)-\mbox{erf}(\pi-\alpha))], \\ \nonumber
 p_{2}^{(+)\theta}& =& \mathcal{A}_{\theta}^{2}\pi \exp(-\pi^{2}).
\end{eqnarray}
The Heisenberg uncertainty product for $\alpha \geq 0$ is then
\begin{eqnarray}\label{}
  \nonumber \Delta_{|m,\alpha,\theta\rangle}\hat{Q}\cdot
  \Delta_{|m,\alpha,\theta\rangle}\hat{P}^{\theta} = \\
  = \sqrt{\frac{1}{2} +p_{2}^{(+)\theta}}\cdot
  \sqrt{\frac{1}{2}+q_{2}^{(+)\theta}(\alpha)-2\alpha
  q_{1}^{(+)\theta}(\alpha)-q_{1}^{(+)\theta}(\alpha)^{2}}.
\end{eqnarray}
The discussion about possible relevance of the uncertainty product
is postponed to the next section.

\section{Conclusion}

This work was devoted to a construction of coherent states on the
circle and investigation of their properties. We used quantizations
on the circle with and without an Aharonov-Bohm type flux with
parameter $\theta$ related to the magnetic flux through the circle.
In these cases we introduced Weyl operators, which were then used to
construct group-related coherent states in the sense of Perelomov.
If the parameter $\theta$ vanishes, then the results of section 5
fully correspond to the results without magnetic flux given in
sections 2--4, as expected.

For the obtained families of coherent states the property of
resolution of unity was proved. Also their overlaps and matrix
elements were expressed using the analytic error function
$\mbox{erf}(z)$. Some results were calculated numerically or
evaluated with the help of MATHEMATICA. For instance, the absolute
value of the inner product is plotted in figures 1--3 for three
choices of the parameters. We have briefly reported on the matter in
\cite{ChLT}. On the one hand, we did not dwell on some evident
consequences of the resolution of unity like the reproducing kernel
property of the overlaps. Also the issue of a Bargmann-Segal
representation seems to require a deeper study because of integral
values of parameter $m$.

On the other hand, we devoted much effort to compare our coherent
states with canonical coherent states which provide wave packets
minimizing Heisenberg's uncertainty relations. For this reason the
circle --- the configuration space --- was identified with the
interval $ [-\pi,\pi)$. The action of $\hat{Q}$ (or $\hat{Q}^{2}$)
considered on $L^{2}(-\pi,\pi)$ is well defined because $\hat{Q}$ is
bounded. However, the momentum operator $\hat{P}$ is unbounded.
Therefore the Heisenberg uncertainty relation is valid only on a
very narrow set of states which belong to a common invariant domain
of self-adjoint operators $\hat{Q}$, $\hat{P}$. It is very clearly
described e.g. in Chapter 8 of \cite{BEH} that such a domain exists
and the Heisenberg uncertainty relation is valid on it (see also
\cite{Kastrup} and the references therein).

However, our coherent states do not belong to this domain, in
particular because they violate conditions at the ends of the
interval $[-\pi,\pi]$. Especially they are not in the domains of
operators $\hat{P}^{2}$ and $\hat{P}\hat{Q}$. Therefore, for the
sake of evaluating the dispersion $\Delta \hat{P}$ and its
comparison with canonical coherent states, we considered $\hat{P}$
and $\hat{P}^{2}$ as formal differential operators. This may explain
our results, notably formula (\ref{Heis}) showing that Heisenberg's
inequality is violated for coherent states with $\alpha$ close to
$0$.

Summarizing, we arrived at limits of similarity between our coherent
states and canonical coherent states. In particular we cannot use
Heisenberg's uncertainty theorem which guarantees the well known
inequality, because our states do not fulfil assumptions of this
theorem. Our coherent states are well defined as elements of the
Hilbert space $\mathcal{H}=L^{2}(S^{1},d\varphi)$, but Heisenberg's
theorem requires to essentially narrow down the set of admissible
states. Note that Heisenberg's theorem cannot be applied even to
eigenstates of $\hat{P}$ or $\hat{P}^2$ since they do not belong to
the domain of $\hat{P}\hat{Q}$.

Let us remind that the Aharonov-Bohm type quantizations of
\cite{DST01} were studied by several alternative methods: see e.g.
\cite{Schulman} (Feynman's path integral in non-simply connected
spaces), \cite{Martin} (self-adjoint extensions of the momentum
operator) and \cite{GMS81} (non-relativistic current algebras). For
a thorough discussion of quantizations on the circle with and
without magnetic flux we refer also to the recent article
\cite{Kastrup}.

\ack The support by the Ministry of Education of Czech Republic
(projects MSM6840770039 and LC06002) is gratefully acknowledged. The
authors are grateful to the referees for constructive remarks which
helped to improve the presentation.

\end{document}